\documentclass[a4paper,twocolumn,11pt]{quantumarticle}
\pdfoutput=1
\usepackage[letterpaper, margin=1in, headheight=15pt]{geometry}
\usepackage[utf8]{inputenc}
\usepackage[english]{babel}
\usepackage[T1]{fontenc}
\usepackage{amsmath}
\RequirePackage[colorlinks=true, allcolors=blue]{hyperref}
\usepackage[braket, qm]{qcircuit}
\usepackage{fancyhdr}
\usepackage{graphicx}
\usepackage{xcolor}
\usepackage{comment}
\usepackage{physics}
\usepackage{float}
\usepackage{hyperref}
\usepackage{svg}

\usepackage{mathtools}
\usepackage{amssymb}
\usepackage{xurl}
\DeclarePairedDelimiter{\ceil}{\lceil}{\rceil}

\begin{document}

\title{Quantum Computing for Optimizing Aircraft Loading}

\author{Ananth Kaushik}
\affiliation{IonQ Inc, 4505 Campus Dr, College Park, MD 20740, USA}
\orcid{0000-0002-2445-2701}
\email{kaushik@ionq.co}
%\homepage{http://quantum-journal.org}
%\thanks{You can use the \texttt{\textbackslash{}email}, \texttt{\textbackslash{}homepage}, and \texttt{\textbackslash{}thanks} commands to add additional information for the preceding \texttt{\textbackslash{}author}. If applicable, this can also be used to indicate that a work has previously been published in conference proceedings.}
\author{Sang Hyub Kim}
\affiliation{IonQ Inc, 4505 Campus Dr, College Park, MD 20740, USA}
\author{Willie Aboumrad}
\affiliation{IonQ Inc, 4505 Campus Dr, College Park, MD 20740, USA}
\author{Martin Roetteler}
\affiliation{IonQ Inc, 4505 Campus Dr, College Park, MD 20740, USA}
\author{Albana Topi}
\affiliation{Airbus Flight Physics Capability Operations, 21129 Hamburg, Germany}
\author{Richard Ashworth}
\affiliation{Airbus Central Research and Technology, AIRTeC bldg 07Y, Bristol BS34 7PA, UK }
\orcid{0000-0003-4446-9559}
\email{richard.ashworth@airbus.com}

%\author{Jasper Krauser}
%\affiliation{Airbus Central Research and Technology, Ottobrunn, Germany}
%\email{jasper.krauser@airbus.com}

\maketitle

\begin{abstract}
   The aircraft loading optimization problem is a computationally hard problem with the best known classical algorithm scaling exponentially with the number of objects. %\textcolor{green}{This would be for a brute force approach exploring all valid and invalid solutions} 
   We propose a quantum approach based on a multi-angle variant of the QAOA algorithm (Multi-Angle Layered Variational Quantum Algorithm (MAL-VQA)) designed to utilize a smaller number of two qubit gates in the quantum circuit as compared to the standard QAOA algorithm so that the quantum optimization algorithm can be run on near-term ion-trap quantum processing units (QPU). We also describe a novel cost function implementation that can handle many different types of inequality constraints without the overhead of introducing slack variables in the quantum circuit so that larger problems with complex constraints may be represented on near-term QPUs which have low qubit counts. We demonstrate the performance of the algorithm on different instances of the aircraft loading problem by execution on IonQ QPUs Aria and Forte. Our experiments obtain the optimal solutions for all the problem instances studied ranging from 12 qubits to 28 qubits. This shows the potential scalability of the method to significantly larger problem sizes with the improvement of quantum hardware in the near future as well as the robustness of the quantum algorithm against varying initial guesses and varying constraints of different problem instances. 
\end{abstract}

\section{Introduction}
Operational efficiency of carrying cargo by air is driven by maximizing revenue-generating aircraft payload and minimizing fuel burn, the main source of costs. This also contributes to meeting sustainability targets. To achieve these objectives, airlines must make the best choice regarding taking cargo onboard while remaining within the constraints of each type of aircraft. At present, aircraft loading choices are often made by airline ground personnel who use their experience and judgment when assigning containerized cargo to aircraft. 

The aircraft loading optimization problem concerns the 
selection of payload containers from among those available and the placement of the containers within the aircraft hold.
%making the best choices for which parts of the available payload to take on board, and where to place them on the aircraft. 
%An airline tries to make best use of the aircraft’s payload capabilities in order to maximise revenue, and to optimise parameters with performance impact towards lower operating costs (fuel burn).
The space for optimization is limited by the operational envelope of the aircraft, which must be respected at all times. Notable limits here are the maximum payload capacity of the aircraft on a specific mission, the center of gravity position of the loaded aircraft and its fuselage shear limits. The aircraft loading problem can be considered as a type of assignment problem - a problem which seeks to find the optimal assignment of cargo to specific locations on the aircraft. The maximum payload capacity constraint imposed on the problem makes it similar to the knapsack problem, a well known NP-Hard optimization problem where the total weight of objects added to the knapsack must not exceed a maximum weight. The knapsack problem is known to be computationally very hard with the best known classical algorithm scaling exponentially with the number of objects making it intractable on a classical computer for a large number of objects. The additional constraints on center of gravity, shear forces on the aircraft etc. do not diminish the complexity of the problem. Quantum computers offer the possibility of better scaling for such problems using the principles of superposition and entanglement. It is therefore, worthwhile exploring quantum algorithms for tackling such classically intractable problems within the NISQ era of quantum computing.

\section{Background}
Combinatorial optimization is the process of finding the minima or maxima of an objective function over a discrete finite set. Problems in operational optimization, such as optimal scheduling, routing, and resource allocation, are the most prominent instances of combinatorial problems encountered in everyday business cases. A key feature of these problems is an exponentially scaling solution space with increasing system size, which makes them particularly hard to tackle using classical algorithms. Most trivially, combinatorial problems can be solved by exhaustive search, that is enumerating every possibility and finding the one that best satisfies solution criteria. However, this search cannot generally be completed in polynomial time making the brute force approach intractable for many practical applications.

Constrained combinatorial problems define a number of conditions that must be satisfied by the solution in addition to minimizing or maximizing the objective function. The constraints have the effect of restricting the solution space, but even with this there is no guarantee that an efficient classical algorithm exists to explore the constrained space. 

These types of problems fall into a complexity class of problems called NP-hard. Problems in this class have exact classical algorithms which only scale exponentially in time complexity, space complexity, or both. There are no known polynomial time scaling classical algorithms for solving these combinatorial problems exactly, although polynomial scaling algorithms may exist for certain special cases or approximations. Some typical examples of problems in this class are max-cut, traveling salesman, knapsack and others.

\section{Aircraft Loading Problem with Constraints}\label{problem_approach}
\subsection{Problem Definition}

\begin{figure}[h]
\centering
\includegraphics[width=0.5\textwidth]{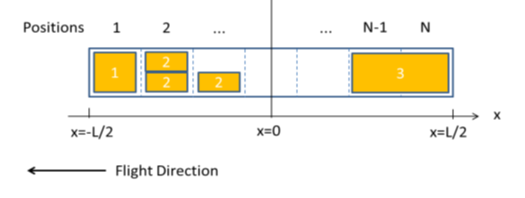}
\caption{The Aircraft Loading Problem: A set of $n$ cargo containers of up to three different sizes is available for loading. Standard size containers (1) occupy a single position, half size containers (2) may share a single position, whereas double size containers (3) occupy two adjacent positions. Each container in this set has an individual mass $m$, which lies in between the empty mass and the maximum mass of each container type. Typically, the combined maximum masses of all containers exceed the aircraft's payload capacity.}
\label{fig:airbus}
\end{figure}

A cargo aircraft is loaded with containers of three standard sizes from a collection of containers available for loading. Each of those containers has a known individual mass. The objective is to maximize the mass of the payload without exceeding a predefined aircraft capacity and while remaining within other constraints, such as aircraft balance (Figure ~\ref{fig:airbus}). 

%\subsection{Problem Definition}

To view this problem as a constrained optimization problem, we define the cost function $H(X)$ to correspond to the total weight of the containers loaded onto the aircraft. The argument $X$ identifies a choice of containers to load and which cargo positions they occupy on the aircraft. Therefore, we seek an arrangement $X_0$ from a set of choices that satisfy all the constraints, such that it maximizes the cost function:
\begin{equation}\label{eqn:max_cost_function}
    X_0 = \mathop{\mathrm{argmax}}_X H(X).
\end{equation}

\begin{figure}[h]
\centering
\includegraphics[width=0.25\textwidth]{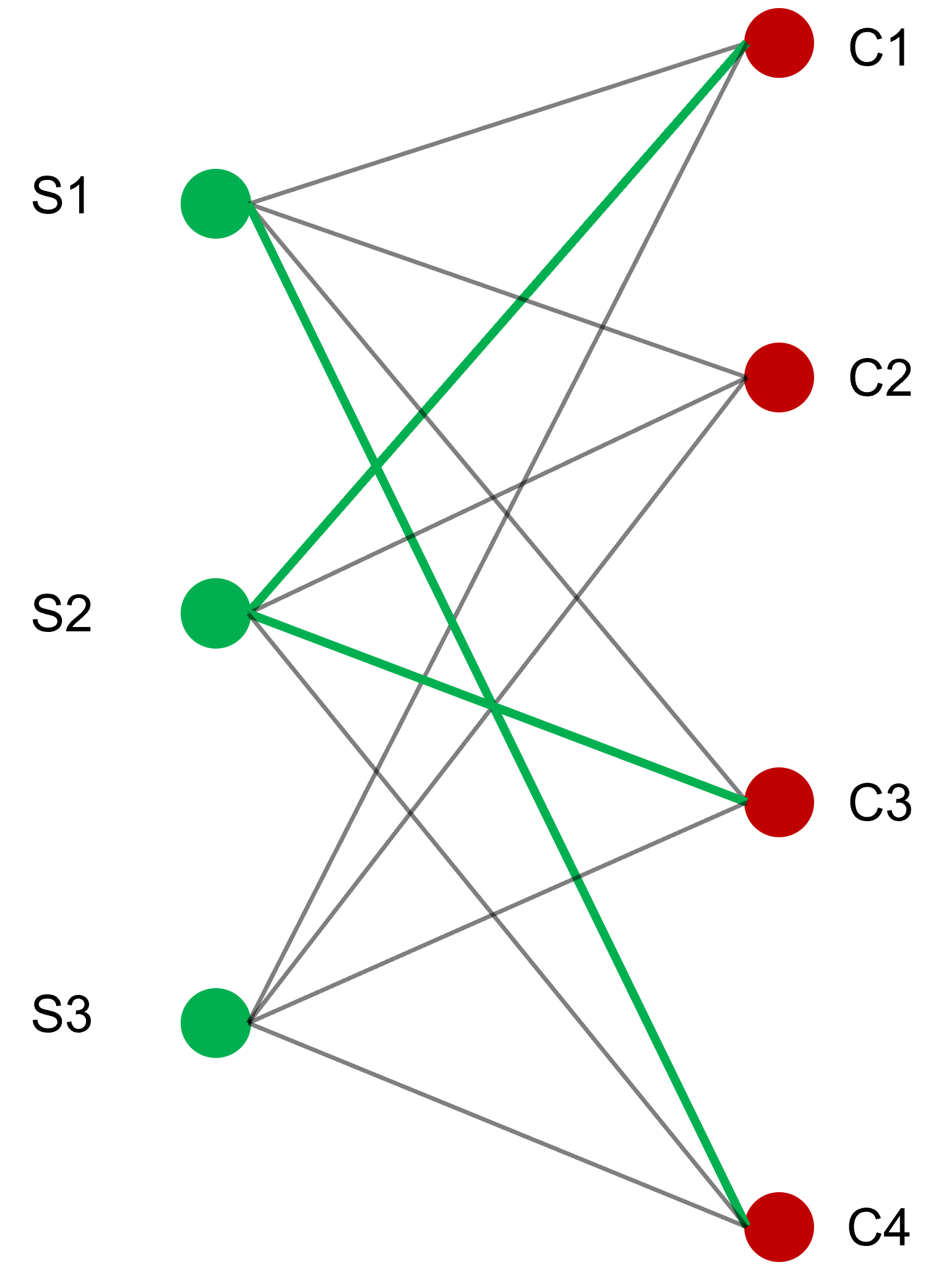}
\caption{Bipartite graph with nodes representing containers (red vertices) and slots (green vertices). Highlighted edges indicate container-to-slot assignments.}
\label{fig:problem_graph}
\end{figure}

As depicted in Figure~\ref{fig:problem_graph}, $X$ can be represented on a bipartite graph with $N$ nodes corresponding to cargo positions on the aircraft and $M$ nodes corresponding to individual containers. Graph edges identify assignments of individual containers to slots. Equivalently, $X$ can be represented as an $M \times N$ matrix  with entries $x_{ij}$ such that $x_{ij} = 1$ if container $i$ is loaded on the aircraft 
in slot $j$, and zero otherwise. With the weights of individual containers given by $w_i$, the total weight of the payload (i.e. the cost function) becomes
\begin{equation}\label{eqn:cost_function}
    H_{obj} = \sum_{i\sim j} w_{i} x_{ij},
\end{equation}
where $\sim$ means all $(i,j)$ pairs of nodes that have an edge defined between them in the graph.

The objective of the optimization problem now becomes the maximization of this total payload weight on the aircraft. Additionally, constraints could be handled as additional penalty terms added to the objective according to the QUBO (quadratic unconstrained binary optimization) formulation 

\begin{equation}\label{eqn:constraints}
    H_{QUBO} = H_{obj} + \sum_{i} P_{i}H_{Ci},
\end{equation}
where $H_{Ci}$ are the Hamiltonians for each of the constraints of the problem and $P_i$ are the penalties applied to the violation of each constraint. However, see below, for practical reasons we opt to handle the constraints outside of the minimization in the quantum circuit.

%This total weight i.e., needs to be maximized. One method of encoding the cost function is using binary variable to encode whether a specific container is loaded or not. Furthermore, this container will need to be assigned a slot on the aircraft. This fact may be encoded using the same binary variable but defined as $x_{ij}$ where $x_{ij} = 1$ if container $i$ is loaded on the aircraft and inserted into slot $j$. Then the cost function becomes: $ H = \sum_{i\sim j} w_{Ci} x_{ij}$ where $w_{Ci}$ is the mass of container $Ci$. 

%With this encoding, it is easy to see that the problem can be represented as a bipartite graph as shown in the figure above. The total number of edges of the graph = number of containers * number of slots

\subsection{Quantum Encoding}

We now adapt the definition of the problem to quantum computing by mapping the cost function onto an Ising Hamiltonian. The binary variable $x_{ij}$ represents the number operator in the Ising Hamiltonian and can be decomposed into Pauli operators:
\begin{equation}\label{eqn:ising_map}
    x_{ij} = \hat{n}_k = \frac{1-\hat{Z}_k}{2}, \mathrm{~~~where~~~}  k = iN+j.
\end{equation}

With this mapping, the total number of qubits required will be equal to the number of edges in the bipartite graph, $MN$, and the $MN$-qubit quantum state is interpreted such that groups of $N$ qubits specify the assignment of a container to zero, one, or two cargo positions, for example for the graph in Figure~\ref{fig:problem_graph}:
$$
|
\underbrace{010}_{C_1} \underbrace{000}_{C_2} \underbrace{010}_{C_3} \underbrace{100}_{C_4}
\rangle
$$

With this encoding strategy, we will now proceed to constructing a parametrized quantum circuit, which represents the cost function. After the optimization has completed and the circuit parameters are found, the most probable measurement outcomes are likely to encode the solutions, which maximize the cost function and do not violate any constraints.

\subsection{Multi-Angle Layered Variational Optimization Algorithm (MAL-VQA)}

Here we propose an alternative \textcolor{blue}{\cite{Herrman2022multi-angle}} to QAOA in order to reduce the circuit depth so that the optimization can be run on a near-term QPU. It has many of the same properties as QAOA, but leads to much shallower circuits since each gate on the quantum circuit has a unique parameter as compared to the standard QAOA where the entire ``Mixer'' block or the ``Hamiltonian'' block which consist or several gates is parametrized by a single parameter. This method is generally known in the literature \textcolor{blue}{\cite{Herrman2022multi-angle}} as the Multi-angle  Quantum Alternating Operator Algorithm (MA-QAOA). 

The presence of inequality constraints in the optimization problem is handled in the QUBO formulation through the introduction of additional penalty terms in the Hamiltonian with associated slack variables as their coefficients. These slack variables also have to be evolved using the quantum circuit requiring extra qubits. The number of extra qubits required scales poorly with  increasing system size and the number of inequality constraints. This also leads to very deep circuits which are difficult to implement in NISQ quantum devices. In this study, we introduce a novel approach where these inequality constraints are instead incorporated as a penalty during the computation of the expectation value alone thereby effectively offloading them to the classical optimizer. The quantum circuit can therefore, be made simpler, more effective and have a flexible architecture. 

\begin{figure}[h]
\centering
\raisebox{2.5cm}{a)}\includegraphics[width=0.45\textwidth]{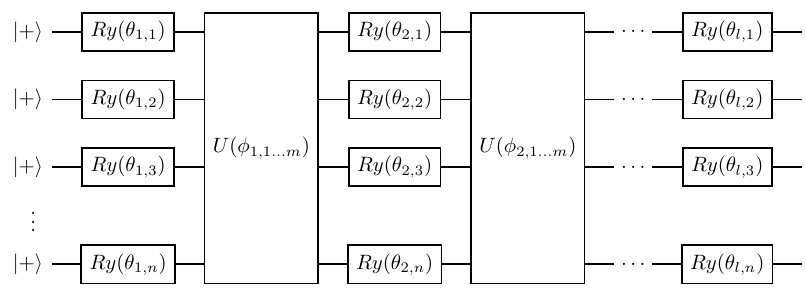}
\hfill
{b)}\raisebox{0.5cm}{\scalebox{0.608}{
\Qcircuit @C=1.0em @R=1.0em @!R { \\
                \nghost{{q}_{0} :  } & \lstick{{q}_{0} :  } & \multigate{1}{\mathrm{R_{ZZ}}\,(\mathrm{{\ensuremath{\theta}}_0})}_<<<{0} & \qw & \qw & \qw & \multigate{4}{\mathrm{R_{ZZ}} (\mathrm{{\ensuremath{\theta}}_4})}_<<<{0} & \qw\\
                \nghost{{q}_{1} :  } & \lstick{{q}_{1} :  } & \ghost{\mathrm{R_{ZZ}}\,(\mathrm{{\ensuremath{\theta}}_0})}_<<<{1} & \multigate{1}{\mathrm{R_{ZZ}}\,(\mathrm{{\ensuremath{\theta}}_1})}_<<<{0} & \qw & \qw & \ghost{\mathrm{R_{ZZ}}\,(\mathrm{{\ensuremath{\theta}}_4})} & \qw\\
                \nghost{{q}_{2} :  } & \lstick{{q}_{2} :  } & \qw & \ghost{\mathrm{R_{ZZ}}\,(\mathrm{{\ensuremath{\theta}}_1})}_<<<{1} & \multigate{1}{\mathrm{R_{ZZ}}\,(\mathrm{{\ensuremath{\theta}}_2})}_<<<{0} & \qw & \ghost{\mathrm{R_{ZZ}}\,(\mathrm{{\ensuremath{\theta}}_4})} & \qw\\
                \nghost{{q}_{3} :  } & \lstick{{q}_{3} :  } & \qw & \qw & \ghost{\mathrm{R_{ZZ}}\,(\mathrm{{\ensuremath{\theta}}_2})}_<<<{1} & \multigate{1}{\mathrm{R_{ZZ}}\,(\mathrm{{\ensuremath{\theta}}_3})}_<<<{0} & \ghost{\mathrm{R_{ZZ}}\,(\mathrm{{\ensuremath{\theta}}_4})} & \qw\\
                \nghost{{q}_{4} :  } & \lstick{{q}_{4} :  } & \qw & \qw & \qw & \ghost{\mathrm{R_{ZZ}}\,(\mathrm{{\ensuremath{\theta}}_3})}_<<<{1} & \ghost{\mathrm{R_{ZZ}}\,(\mathrm{{\ensuremath{\theta}}_4})}_<<<{1} & \qw\\
\\ }}}
\caption{Architecture of the quantum circuit used in the study. a) The quantum circuit consists of alternating layers of parametrized single qubit rotation $R_Y$ gates and entanglement blocks U. The U blocks consist of parametrized $R_{ZZ}$ gates between different qubits. In this study, only one entanglement block U was used sandwiched between two layers of $R_Y$ gates, b) The arrangement of $R_{ZZ}$ gates within a single block U used in this study shown for a 5 qubit circuit.}
\label{fig:qnn_arch2}
\end{figure}

The quantum circuit architecture used in this study is shown in Fig~\ref{fig:qnn_arch2}(a). The quantum circuit consists of alternating layers of parametrized single qubit rotation $R_Y$ gates and entanglement unitaries U similar to the standard QAOA structure. The single qubit $R_Y$ gates serve as a simple mixer for the quantum states. The key difference from QAOA is that each $R_Y$ gate can have a different parameter as opposed to a single parameter per layer in standard QAOA. The entanglement unitaries U entangle multiple qubits and may be constructed with different types of entangling gates and can have any entanglement structure. Note that this is in contrast to standard QAOA where the entanglement unitary block must exponentiate the Hamiltonian associated with the problem. The flexibility of having a unique parameter per gate and a flexibile design of the entanglement unitaries enable very short circuit depths while retaining expressibility of the ansatz to represent the desired Hilbert space of solutions to the optimization problem. In this study, we used 2 qubit $R_{ZZ}$ gates to define the entanglement unitaries U with a ladder like structure as shown in Fig~\ref{fig:qnn_arch2}(b). 

There are, however, a few potential disadvantages to this approach when compared to the standard QAOA. There is no formal guarantee that the quantum algorithm will find the ground state of the Hamiltonian even as the number of layers $p\to\infty$. The ability to find the ground state will depend on the type of optimizer used, the form of the chosen ansatz, number of parameters in the ansatz etc. Also, unlike QAOA, adding more layers of the ansatz does not necessarily improve the ability of the algorithm to find the ground state - i.e. there is no straightforward mapping to the adiabatic theorem.

\subsubsection{Incorporating Constraints}
\paragraph{Maximum loading weight}
The total mass of loaded containers may not exceed the payload capacity of the aircraft $W_{max}$:
$$ \sum_{i \sim j} w_{i} x_{ij} \leq W_{max}. $$
In the conventional QAOA approach, this inequality constraint is encoded into the Hamiltonian  requiring extra slack qubits. This is not economical in the current era with a limited number of qubits. Therefore, we pursue an alternative method to encode the constraint which adds a penalty if the state produced by the quantum circuit after measurement corresponds to a container assignment that exceeds $W_{max}$. This can be done on a classical computer thereby offloading the constraint evaluation from the quantum computer. This approach saves both the number of qubits needed, as well as the number of quantum gates required.

\paragraph{Center of gravity}
The center of gravity is constrained between $R_{min}$ and $R_{max}$, the lower and upper bounds
with:
$$ R_{min} \leq \frac{\sum_{i \sim j} d_{i}w_{i} x_{ij}}{\sum_{i \sim j} w_{i} x_{ij}} \leq R_{max}, $$
where $d_i$ is the distance of slot $i$ from the center of the aircraft.
Similar to
the maximum loading weight
constraint, this is another inequality constraint, which will require extra slack qubits if encoded into the Hamiltonian. Therefore, we use the penalty approach to represent this constraint too.

\paragraph{Maximum Shear}
For a maximum permitted shear profile $\tau$, given as a function of the distance from the center of the aircraft, we impose the discretized shear constraints

\begin{align*}
\sum_i \bigg(
\frac{1}{2} w_i x_{ij} + \sum_{\ell \in \mathcal{F}_j} w_i x_{i \ell}
\bigg)
\leq \tau(d_j), \\
\quad \text{for each} \quad j = 1, \ldots, m.
\end{align*}

Here $\mathcal{F}_j = \{1, \ldots, j-1\}$ if $d_j < 0$ and $\mathcal{F}_j = \{j + 1, \ldots, m\}$ counts the set of slots that are further from the center than the $j$th slot.

The imposed maximum shear profile can be linear or non-linear. This is another inequality constraint that is handled 
by adding a penalty term to the cost function.

\begin{figure}[h]
\centering
\includegraphics[width=0.45\textwidth]{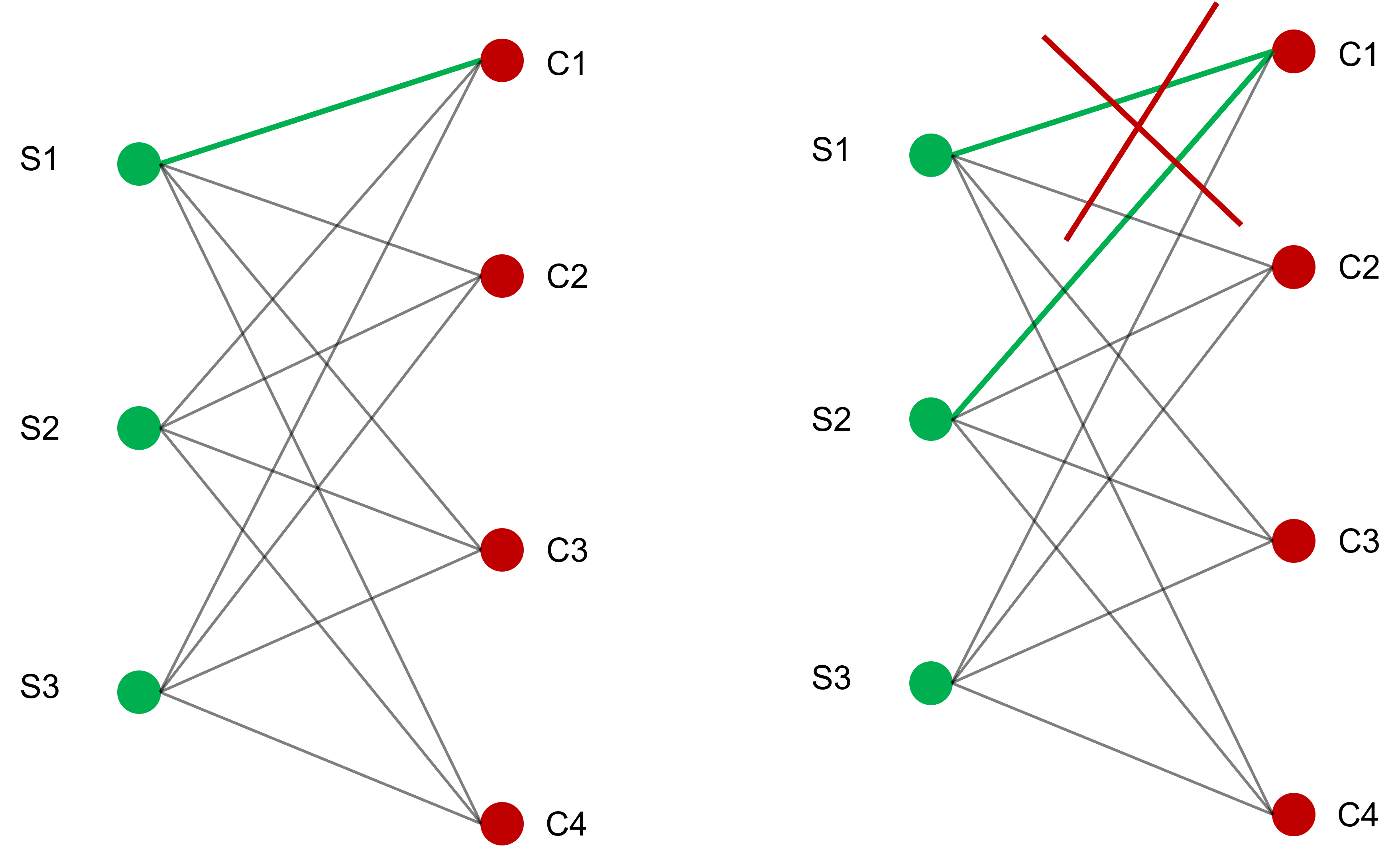}
\caption{Only 1 edge can be selected (shown in green, left image). Two selected edges from the same container not allowed (shown in green with a red cross, right image)}
\label{fig:edge_constraint}
\end{figure}

\paragraph{Slot assignment}
Once a given container of type 1 or type 2 is assigned to a slot, it cannot be assigned to any other slot (Figure~\ref{fig:edge_constraint}). This is an implicit constraint required to prevent a specific container being assigned 
to more than one slot in the solution bitstring. If the selected container is type 3, it needs to be assigned to two consecutive slots in the solution bitstring, and no more than two slots. The constraint is imposed through the equation

\begin{align*}
\sum_{j = 1}^m x_{ij} \leq v_i, 
\quad \text{for each} \quad i = 1, \ldots, n,
\end{align*}

where $v_i = 1$ if $t_i = 1, 2$ and $v_i = 2$ if $t_i = 3$. 

\begin{comment}
\begin{figure}[h]
\centering
\includegraphics[width=0.25\textwidth]{Figs/graph-assignment-options.png}
\caption{Bipartite graph showing edges arising from container C1 are shown in blue.}
\label{fig:C1_to_slots}
\end{figure}
\end{comment}

In addition, we impose the following quadratic constraints to enforce the restriction that type $t_i = 3$ containers may only be assigned to contiguous slots

\begin{align*}
2 \sum_{j=1}^{m-1} x_{i, j} x_{i, j+1} = \sum_{j=1}^m x_{i, j}, \quad \\
\text{for each } i \text{ such that } t_i = 3.
\end{align*}

In order to incorporate these constraints, we use a generic superposition of all possible quantum states which will include those states where a single container has been assigned 
to multiple slots simultaneously. These states are then suppressed  
by adding a penalty term to the total cost function.

\paragraph{Container assignment}
Containers assigned to a given slot should fit within the slot:
$$ \sum_C t_C x_C \leq 1, $$
where 
the sum is over containers and $t_C$ is the container type and $x_C$ is the binary variable
$$
\left\{\begin{array}{ll}
     t_C = 1.0, & \mathrm{type~1} \\
     t_C = 0.5, & \mathrm{type~2} \\
     t_C = 1.0, & \mathrm{type~3}
\end{array}\right.
$$

The constraint is imposed through the equation

\begin{align*}
\sum_{i = 1}^n x_{ij} \leq k_j, 
\quad \text{for each} \quad j = 1, \ldots, m
\end{align*}

where $k_j = 1$ if $t_j = 1, 3$ and $k_j = 1/2$ if $t_j = 2$.

This is another inequality constraint that is handled using a penalty term to the cost function.

%This is an inequality constraint which will require extra slack qubits to encode into the Hamiltonian itself. This would be the traditional manner to encode the constraint if using QAOA. This method does not scale well with number of containers and will be very difficult to implement in the current era with limited number of qubits. An alternative proposal using the QVO route would be to encode the constraint as a penalty while computing the expectation value of states produced by the quantum circuit after measurement if the above inequality is violated. This can be done on a classical computer thereby offloading the constraint evaluation from the quantum computer. This approach will save both on the number of qubits needed as well as the number of quantum gates required.

\paragraph{Prominence of the zero state}
The $|000\dots\rangle$ state (all qubits are in the $|0\rangle$ state) is always a solution to the constraints, so it forms a local minimum. It is possible for the optimizer to get stuck here thereby amplifying the probability of the state. We want to prevent the appearance of the $|000\dots\rangle$ state but preserve all other possible states. This special state is penalized separately while adding the expectation value to the total cost function. 

\subsubsection{The Cost Function}

\begin{figure}[h]
\centering
\includegraphics[width=0.5\textwidth]{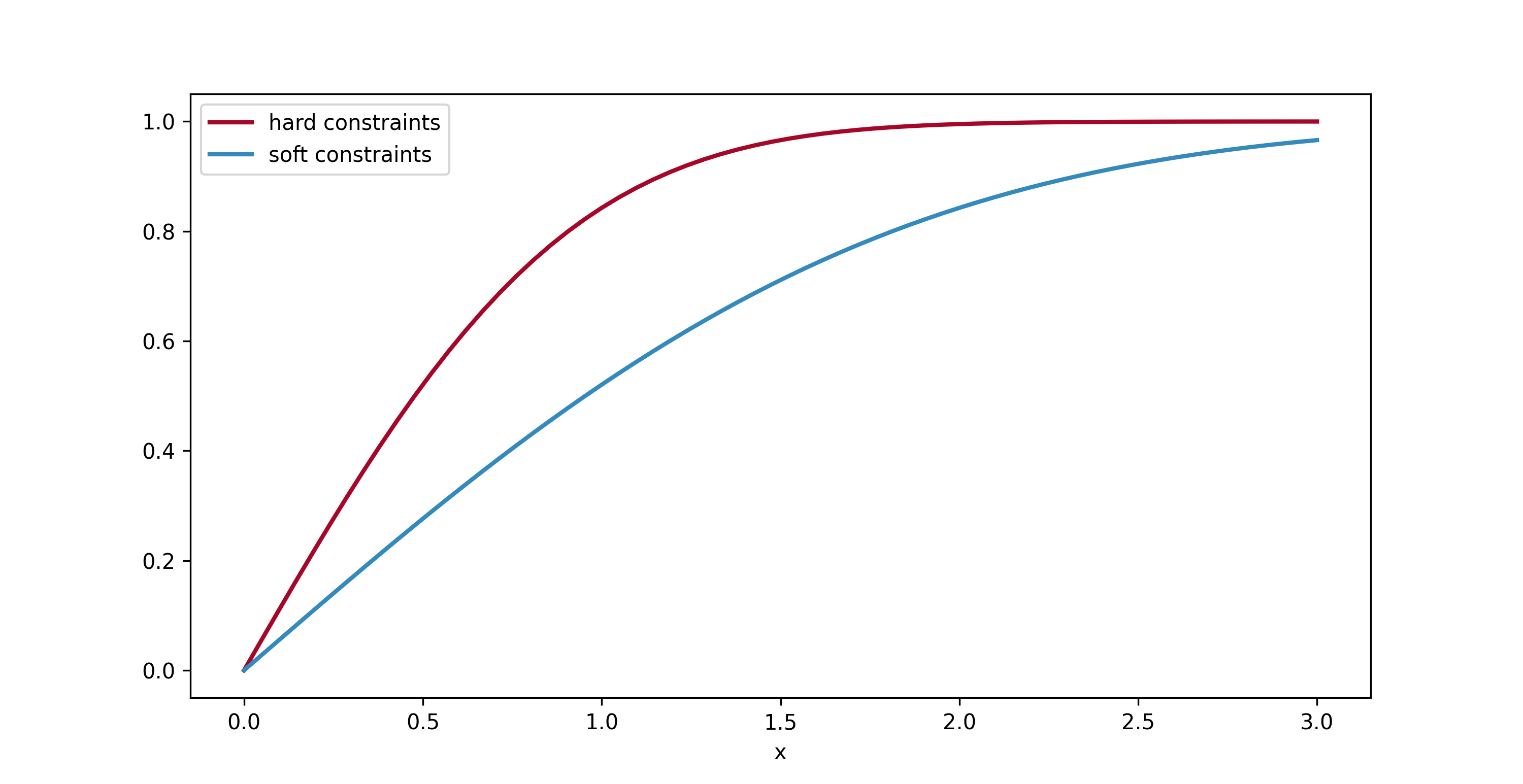}
\caption{Cost function behavior for constraints. The error function multiplied by a penalty is used to implement the constraint. It is approximately linear for small violations of the constraint and gradually saturates at the value of the penalty. The soft constraints are modelled with a softer error function than the hard constraints.}
\label{fig:cf_constraint}
\end{figure}

The constraints defined in the previous section are typically implemented using a quadratic penalty function. These penalties increase the value of the cost function for states that violate the constraints thereby guiding the optimizer to lower the cost by choosing states that satisfy constraints. The penalties are defined to be proportional to the square of the degree of violation of the constraints. The total cost is the expectation value over all component states evaluated on the cost function. 

\begin{align*} 
\text{Expectation} = \sum_{i} \bra{\psi_{i}}H\ket{\psi_{i}} = \sum_{i} p_{i} \cdot cost(\ket{\psi_{i}}) \label{eq:expectation_value}
\end{align*}

The expectation value is computed as a sum of
the product of the probability of a state and the cost of that state. For the states that violate the constraints to a high degree, the cost will be large. If such states have a high probability, then the expectation value would also be large. The optimizer attempts to lower the expectation value during the optimization. The optimizer may, therefore, reduce the total expectation value simply by redistributing the probabilities p$_i$ over states that violate the constraints to a lesser degree and 
by never amplifying the probabilities of the states that do not
violate the constraints. This is 
enabled by the fact that a quadratic penalty increases very slowly allowing the optimizer to make this redistribution of probabilities. 

However, this is not an effective cost function definition since the cost function should enable the optimizer to seek optimal states efficiently. Since the cost function is being evaluated on the CPU, we are free to define any nonlinear function as the the penalty. 

After trials with various non-linear functions an error function was selected for the penalty term. This function rises more steeply than the quadratic function and penalizes the states that violate the constraints appropriately. 

The constraints are grouped into 2 sets - hard constraints and soft constraints. The hard constraints include 
those for the maximum weight, the total shear stress and volume/space constraints for the containers in the slots. The soft constraints include the center of gravity limits. 

The penalty for states that violate the constraints is encoded as an error function multiplied by a suitable penalty value coefficient. The coefficient of the argument of the error function
is also larger for the hard constraints than the soft constraints making the slope much steeper for the hard constraints as seen in Fig~\ref{fig:cf_constraint}. The error function is linear for small violations of the constraints but rises sharply and saturates to the penalty value for larger violations. 
This prevents the optimizer from lowering the total expectation value by shifting probabilities between states that violate constraints to varying degrees since all such states are penalized equally. But the error function is a smooth function that is differentiable so this helps prevent the occurrence of barren plateaus (as opposed to say, a step function which also saturates but would be flat and thus gives rise to barren plateaus). 

Instead of the conventional energy expectation
value, we use the conditional value at risk method (CVaR) as a cost function. The motivation for this approach is to improve the overlap of the final state with the optimal state at the cost of potentially having a higher expectation value. Given a random variable X with cumulative density function $F_X$, the CVaR is defined as the conditional expectation over the left $\epsilon$-tail of the distribution

\begin{equation}
    CVaR_{\epsilon}(X) = \mathbb{E} [X|X \leq F_X^{-1}(\epsilon)]
\end{equation}

In the expression above, $\mathbb{E}$ denotes the expected value, $F_X^{-1}$ the inverse of the cumulative density function, and $\epsilon$ is a parameter in the interval $(0, 1]$. This corresponds to only considering a fraction $\epsilon$ of all measurements with the lowest energies for a given set of variational parameters. More specifically, assuming that we perform K measurements resulting in K computational basis states with corresponding energy values $\{E_1,E_2,\ldots,E_K\}$ sorted in ascending order, the cost function for a given set of parameters is given by

\begin{equation}
    CVaR_{\epsilon} = \frac{1}{\ceil{\epsilon K}} \sum_{i=1}^{\ceil{\epsilon K}} E_i.
\end{equation}

For $\epsilon$=1 the expression above is nothing but the usual estimate of the energy expectation value with K measurements. The limit $\epsilon \to 0$ corresponds to just keeping (one of) the measurement(s) with the lowest energy value, which would lead to a cost function that is discontinuous in the variational parameters. Note that the choice of $\epsilon$ also affects the maximum component of the optimal solution one can expect in the ground state. Since only the fraction $\epsilon$ of measurements with the lowest energy contributes to the cost function, there is essentially no reward for increasing the component of the optimal solution in the final state above $\epsilon$. Thus, in practice one has to choose a reasonable value for $\epsilon$ which allows for generating a sufficiently large component of the optimal solution, but which is small enough to take advantage of the CVaR. The details of the methodology is described in greater detail in \cite{Barkoutsos2020improving}.

\section{Results}\label{results}

Several instances of the aircraft loading problem of different sizes (number of containers and slots)  were executed on IonQ Aria and IonQ Forte trapped ion quantum processing units (QPU). The IonQ Aria QPU uses up to 25 addressable Ytterbium (Yb) ions linearly arranged in an ion trap, while the IonQ Forte QPU has 36 addressable Ytterbium (Yb) ions. Qubit states are implemented by utilizing two states in the ground hyperfine manifold of the Yb ions. Manipulating the qubits in the Aria/Forte QPU is done by 355-nm laser pulses, which drive Raman transitions between the qubit states. By configuring these pulses, arbitrary single-qubit gates and Mølmer-Sørenson type two-qubit gates \cite{sorensen1999quantum} can both be realized. As of 2024, the Aria QPU has demonstrated performance at the level of 25 algorithmic qubits and the Forte QPU has demonstrated performance at the level of 36 algorithmic qubits \cite{IonQ_blog, IonQ_Aria, IonQ_Forte, lubinski2023application}.

In order to mitigate the effect of systematic errors on the Aria/Forte QPU, error mitigation via symmetrization is used \cite{maksymov2023enhancing}. After executing multiple circuit variants with distinct qubit to ion mappings, the measurement statistics is aggregated using component-wise averaging. For all results presented here, we use the CVaR as an aggregation function. To show that current quantum devices are suitable for addressing the aircraft loading problem, we proceed in two steps. First, we demonstrate the full MAL-VQA optimization for a few instances of the problem, where we run the feedback loop between the quantum device and the classical optimizer until convergence. Second, for other problem instances we restrict ourselves to performing inference of the result for the optimal parameters (obtained from optimizing the parameters on the classical simulator) on the quantum hardware. We demonstrate the execution of problem instances of up to 7 containers and 4 slots of the aircraft loading problem which requires 28 binary variables (28 qubits) in the QUBO formulation of the problem.

\subsection{Aria QPU Results}\label{aria_results}

\paragraph{Optimization runs}

\begin{figure}[h]
\centering
{a)}\includegraphics[width=0.5\textwidth]{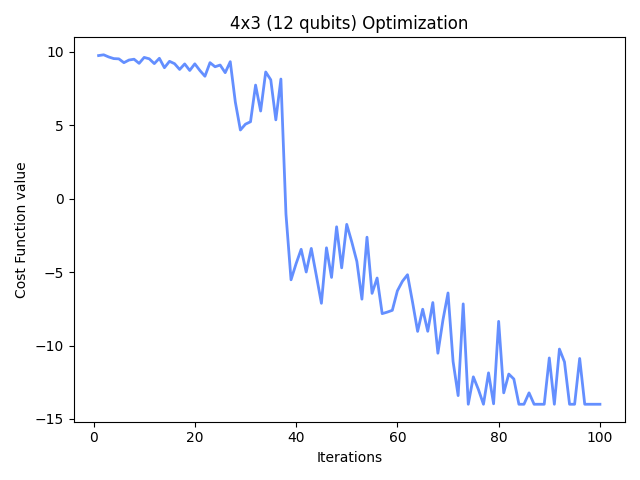}
\hfill
{b)}\includegraphics[width=0.5\textwidth]{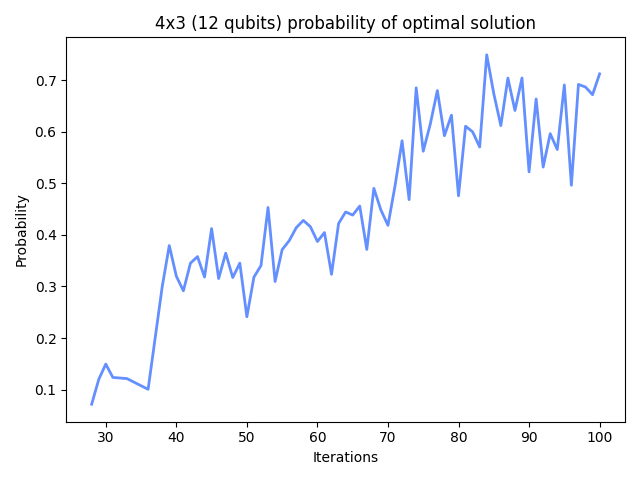}
\caption{Aria QPU results for 4 containers, 3 slots (12 qubits). Maximum loading weight = 14kG. a) The evolution of the cost function during optimization, b) Probability of the optimal solution evolving through the optimization. }
\label{fig:aria_results_4-3_14_optimization}
\end{figure}

The first problem instance chosen for execution of a full hybrid optimization run on IonQ Aria involved 4 containers, 3 slots which requires 12 qubits. The max-weight constraint was selected to be 14 kG. The ansatz structure described in Figure~\ref{fig:qnn_arch2} was chosen with a total of 60 single qubit rotation gates and 24 two qubit entanglement gates. For this problem, 1000 shots were used to sample the output probability distribution and the COBYLA optimizer was used to perform the optimization. A maximum of 100 iterations were used to run the optimization. The evolution of the cost function and the probability of the optimal solution are shown in Fig~\ref{fig:aria_results_4-3_14_optimization}. As the cost function decreases with increasing number of iterations, the probability of the optimal solution gets amplified showing that the optimization begins to converge on the optimal solution. 

\begin{figure}[h]
\centering
{a)}\includegraphics[width=0.3\textwidth]{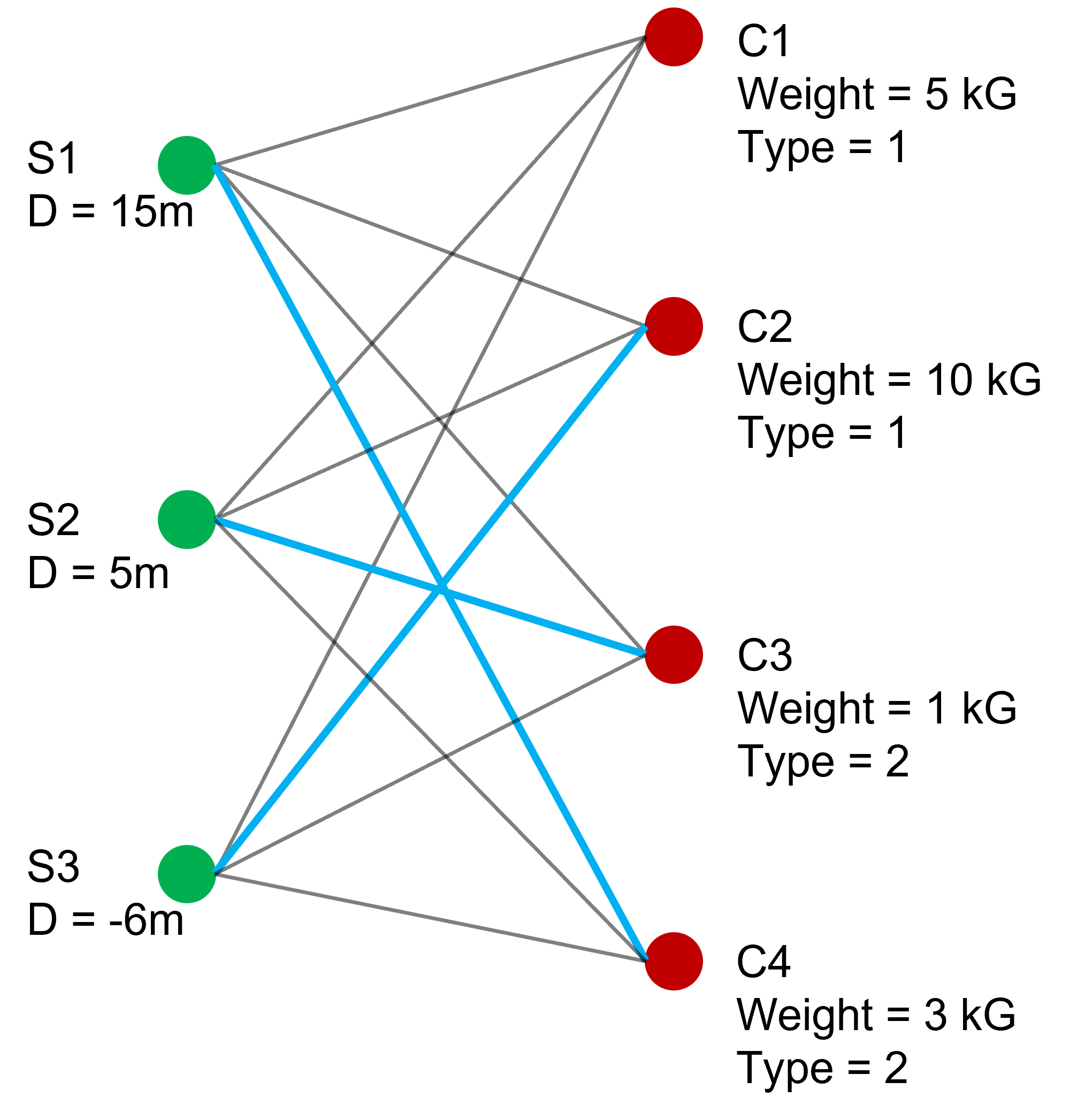}
\hfill
{b)}\includegraphics[width=0.45\textwidth]{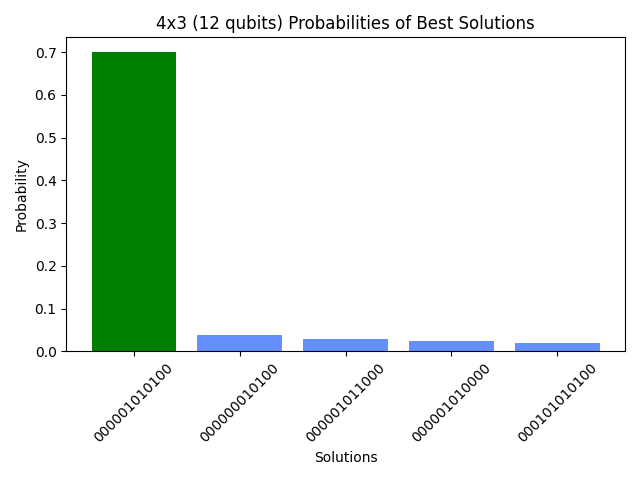}
\caption{Aria QPU results for 4 containers, 3 slots (12 qubits). Maximum loading weight = 14kG. a) The optimal solution for container assignment, b) Probability histogram of results - the green bar corresponds to the optimal solution. }
\label{fig:aria_results_4-3_14_output}
\end{figure}

The probability histogram as well as the optimal solution for the problem are shown in Fig~\ref{fig:aria_results_4-3_14_output}. Only the top 5 highest probability solutions are plotted in the histogram. The results show the optimal solution found with a green bar. The final probability achieved for the optimal solution $\approx$ 70\%. Also noteworthy is that the second highest probability solution is well separated from the optimal solution showing that the probability of the optimal solution is indeed well amplified and can be measured easily. This demonstrates that the Aria QPU has high 2 qubit gate fidelities and low State Preparation and Measurement (SPAM) errors since these are sources of noise inherent in any QPU. 

\paragraph{Inference runs}

In addition to the full hybrid optimization run described in the previous paragraph, two different problem instances with higher qubit counts were executed on the IonQ Aria QPU with pre-optimized variational parameters for the ansatz obtained from running the optimization problem on a classical CPU. The selected problem instances vary in the number of qubits, circuit depth and problem constraints. The details of the selected problem instances are shown below:
\begin{enumerate}
    \item 4 containers, 4 slots: 16 qubits -- max-weight=16 kG -- 1Q gates: 144, 2Q gates: 64
    \item 5 containers, 4 slots: 20 qubits -- max-weight=16 kG -- 1Q gates: 180, 2Q gates: 80
\end{enumerate}

\begin{figure}[h]
\centering
{a)}\includegraphics[width=0.3\textwidth]{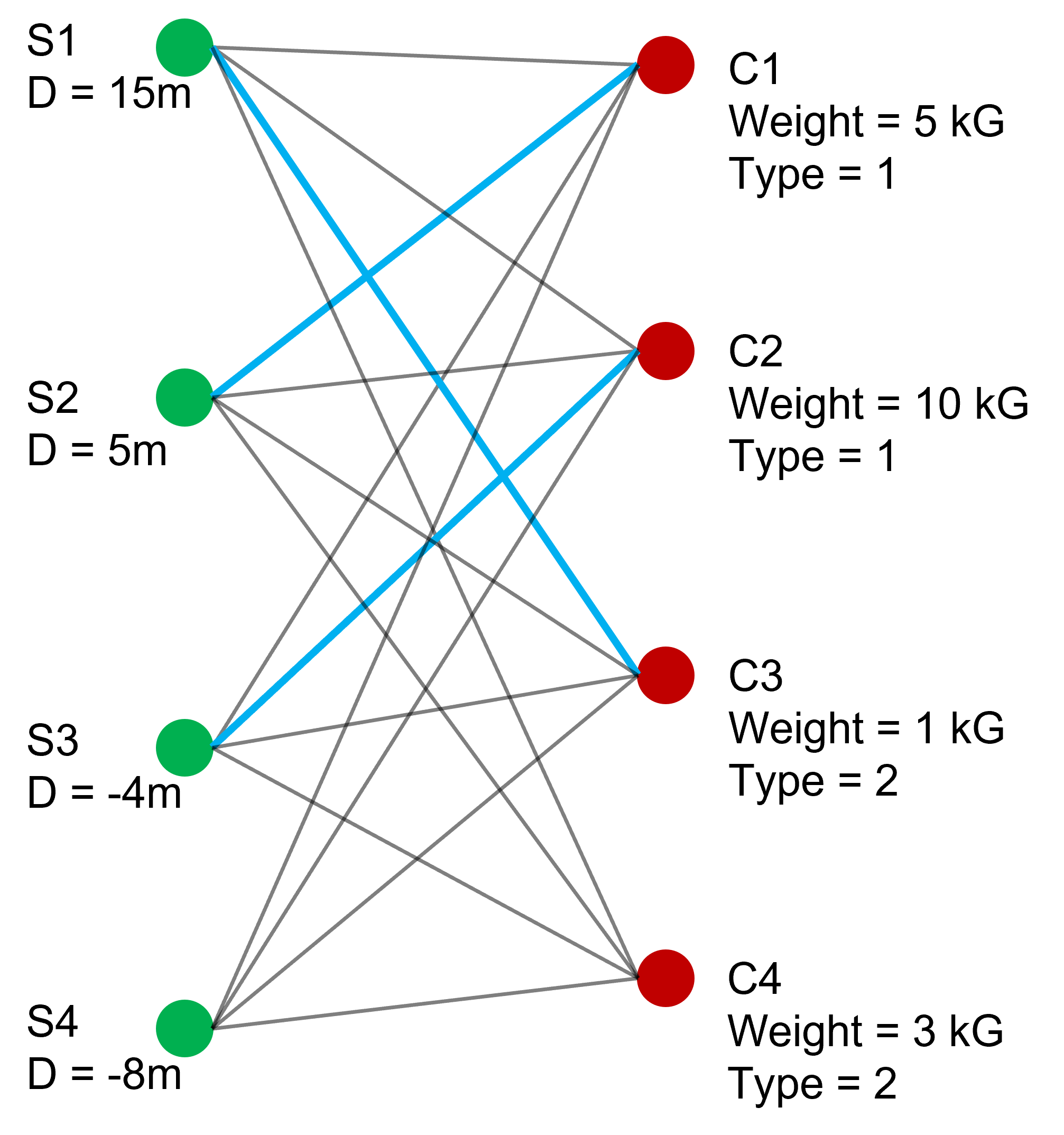}
\hfill
{b)}\includegraphics[width=0.45\textwidth]{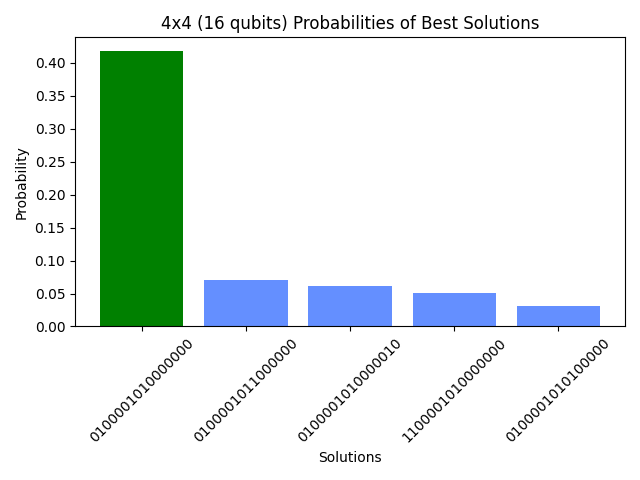}
\caption{Aria QPU results for 4 containers, 4 slots (16 qubits). Maximum loading weight = 16kG. a) The optimal solution for container assignment, b) Probability histogram of results - the green bar corresponds to the optimal solution. }
\label{fig:aria_results_4-4}
\end{figure}

\begin{figure}[h]
\centering
{a)}\includegraphics[width=0.3\textwidth]{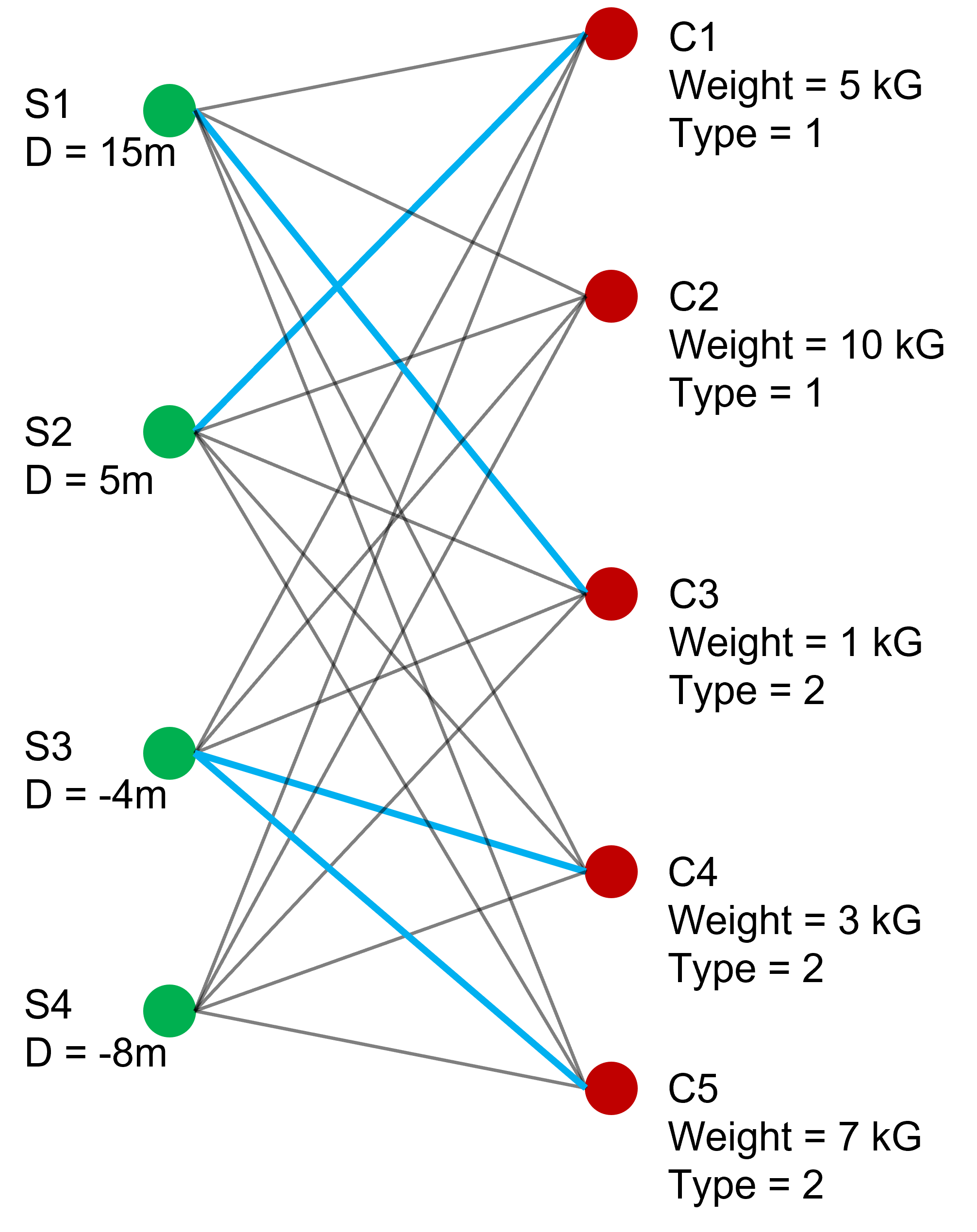}
\hfill
{b)}\includegraphics[width=0.45\textwidth]{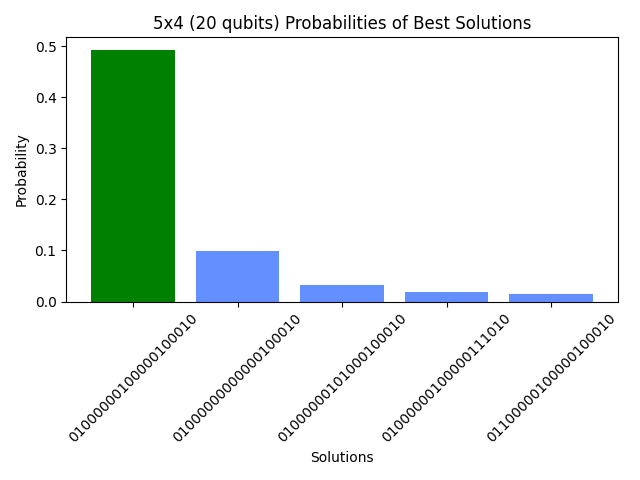}
\caption{Aria QPU results for 5 containers, 4 slots (20 qubits). Maximum loading weight = 16kG. a) The optimal solution for container assignment, b) Probability histogram of results - the green bar corresponds to the optimal solution. }
\label{fig:aria_results_5-4}
\end{figure}

These problem instances were optimized using the qiskit Aer simulator and the final optimal parameters were determined. The optimal parameters were used to initialize the quantum circuit which was then executed on the Aria QPU and the output probability histogram was measured. For each of the problem instances 10000 shots were used to sample the output probability distribution. The probability histograms as well as the optimal solution sampled for the selected problem instances are shown in Figs~\ref{fig:aria_results_4-4} -~\ref{fig:aria_results_5-4}. Only the top 5 highest probability solutions are plotted in the histograms. The results show the optimal solution found in each case with a green bar.

The results show that the optimal solution is the state with the highest probability. The probability of the optimal solution is $\approx$ 40\% for 16 qubit problem instance and $\approx$ 50\% for the 20 qubit problem instance and is well separated from the other sampled solutions showing that the measured probability distribution is not affected greatly by QPU noise. This is true for both problem instances which differ in the number of 2 qubit gates in the quantum circuit and are therefore, affected by QPU noise to different degrees. This demonstrates once again, that the Aria QPU has high 2 qubit gate fidelities and low State Preparation and Measurement (SPAM) errors.

\subsection{Forte QPU Results}\label{forte_results}

\paragraph{Full Optimization}

\begin{figure}[h]
\centering
{a)}\includegraphics[width=0.45\textwidth]{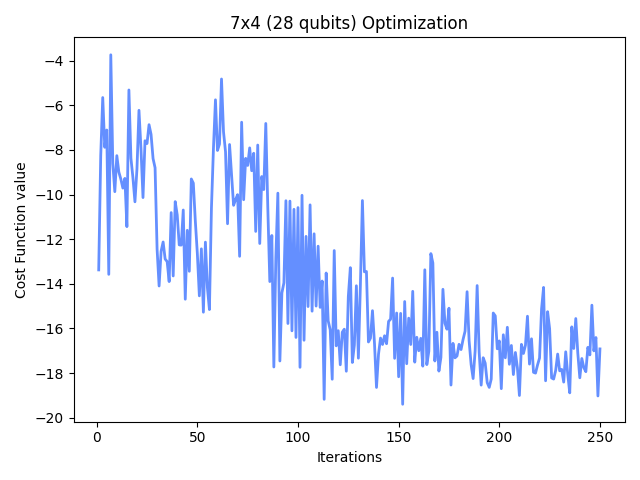}
\hfill
{b)}\includegraphics[width=0.45\textwidth]{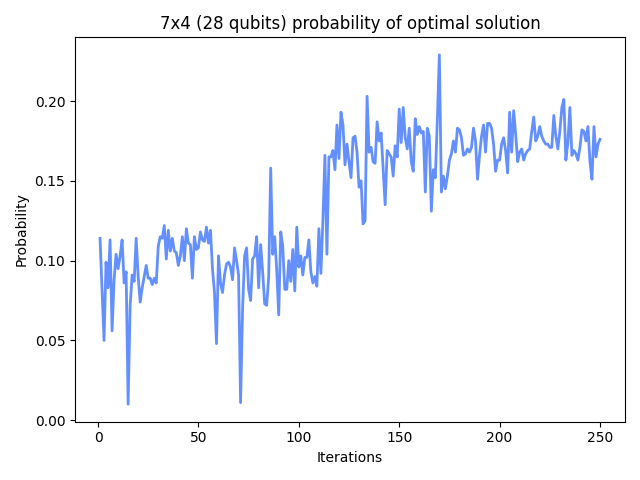}
\caption{Forte QPU full optimization results for 7 containers, 4 slots (28 qubits). Maximum loading weight = 23kG. a) The cost function value as a function of iterations of the optimizer. b) Probability of the optimal solution evolving through the optimization.}
\label{fig:forte_optimization_7-4}
\end{figure}

\begin{figure}[h]
\centering
{a)}\includegraphics[width=0.3\textwidth]{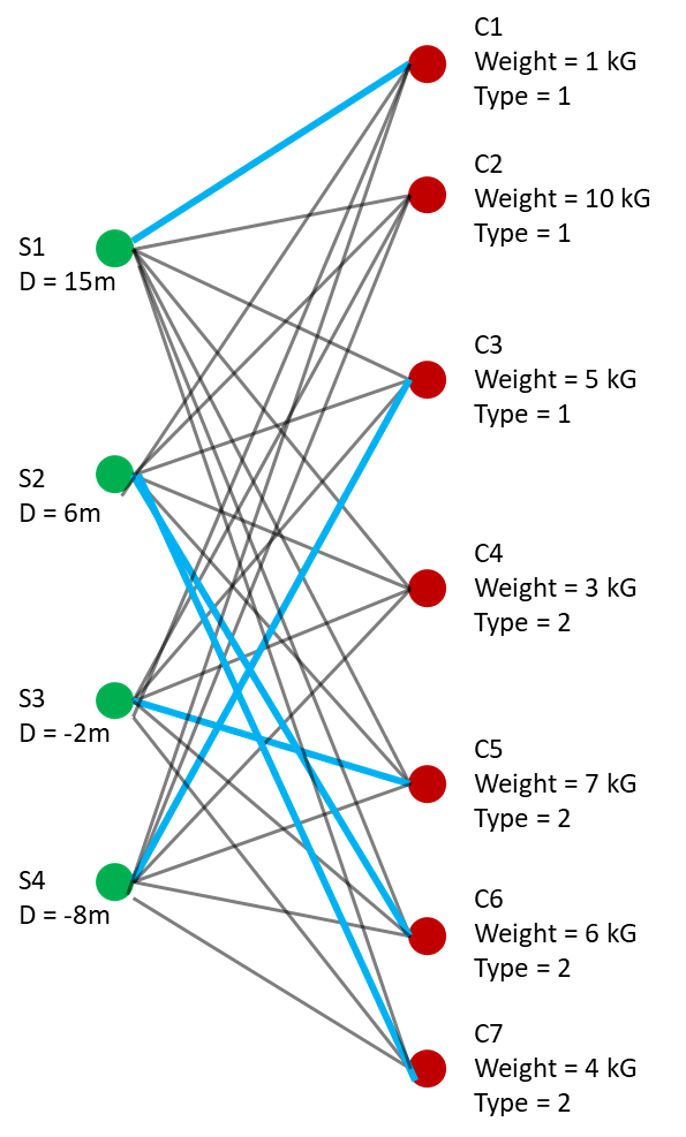}
\hfill
{b)}\includegraphics[width=0.45\textwidth]{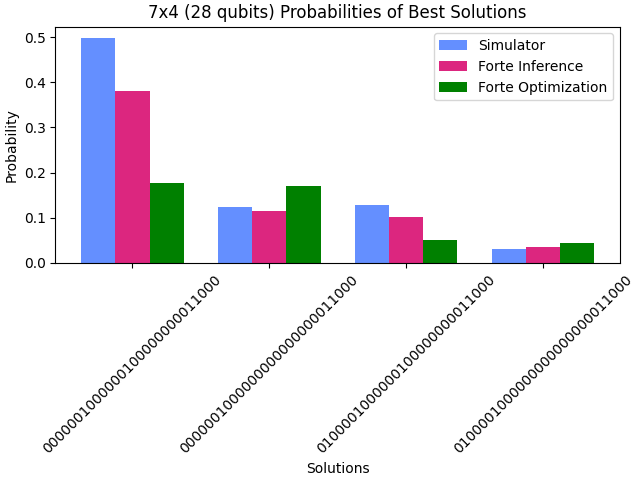}
\caption{Forte QPU results for 7 containers, 4 slots (28 qubits). Maximum loading weight = 23kG. a) The optimal solution for container assignment, b) Probability histogram of results - the first bar on the left corresponds to the optimal solution of total loading weight = 23kG. }
\label{fig:forte_optimization_7-4_results}
\end{figure}

The problem shown in Fig~\ref{fig:forte_optimization_7-4_results}a) was selected to run on the Forte QPU for full hybrid optimization. In order to speed up convergence, the optimization was warm started using a partially converged optimization run on the simulator. Variational parameters from the optimization run on the simulator before the simulation had converged were selected as the initial guess to the optimization run on the Forte QPU. The optimization was then run using the COBYLA optimizer for 250 iterations using 1000 shots per iteration to sample the output probability distribution after measurement of the quantum circuit. The cost function value as a function of the iteration count is shown in Fig~\ref{fig:forte_optimization_7-4}a). The cost function starts at a negative value instead of a positive value as seen in Fig~\ref{fig:aria_results_4-3_14_optimization} because the optimization has been warm started. However, the cost function quickly converges to a much lower value as the optimization proceeds on the Forte QPU. The expectation value is noisy because of the shot noise as well as the QPU noise, but the optimization is still able to converge.  

The final measured probability histogram from the full hybrid optimization experiment is compared against the ideal simulator and the measured histogram from the ``inference'' experiment in 
%Fig~\ref{fig:forte_optimization_7-4}b
Fig~\ref{fig:forte_optimization_7-4_results}b. The top four highest probability solutions are shown. The optimal solution is represented by the first set of bars on the left of the plot. The probability of the optimal solution from the full hybrid optimization experiment on the Forte QPU is lower than the ideal simulator or the ``inference'' experiment due to a combination of shot noise and QPU noise. Nevertheless, the solution is easily distinguishable along with 2 other valid sub-optimal solutions (the 2nd and 4th set of bars, the 3rd set of bars were not valid because they violated the constraints of the problem). These results are a testament to the high 2 qubit gate fidelities of the Forte QPU as well as the robustness of the chosen quantum circuit architecture and cost function design. 

\paragraph{Inference runs}

\begin{figure}[h]
\centering
{a)}\includegraphics[width=0.3\textwidth]{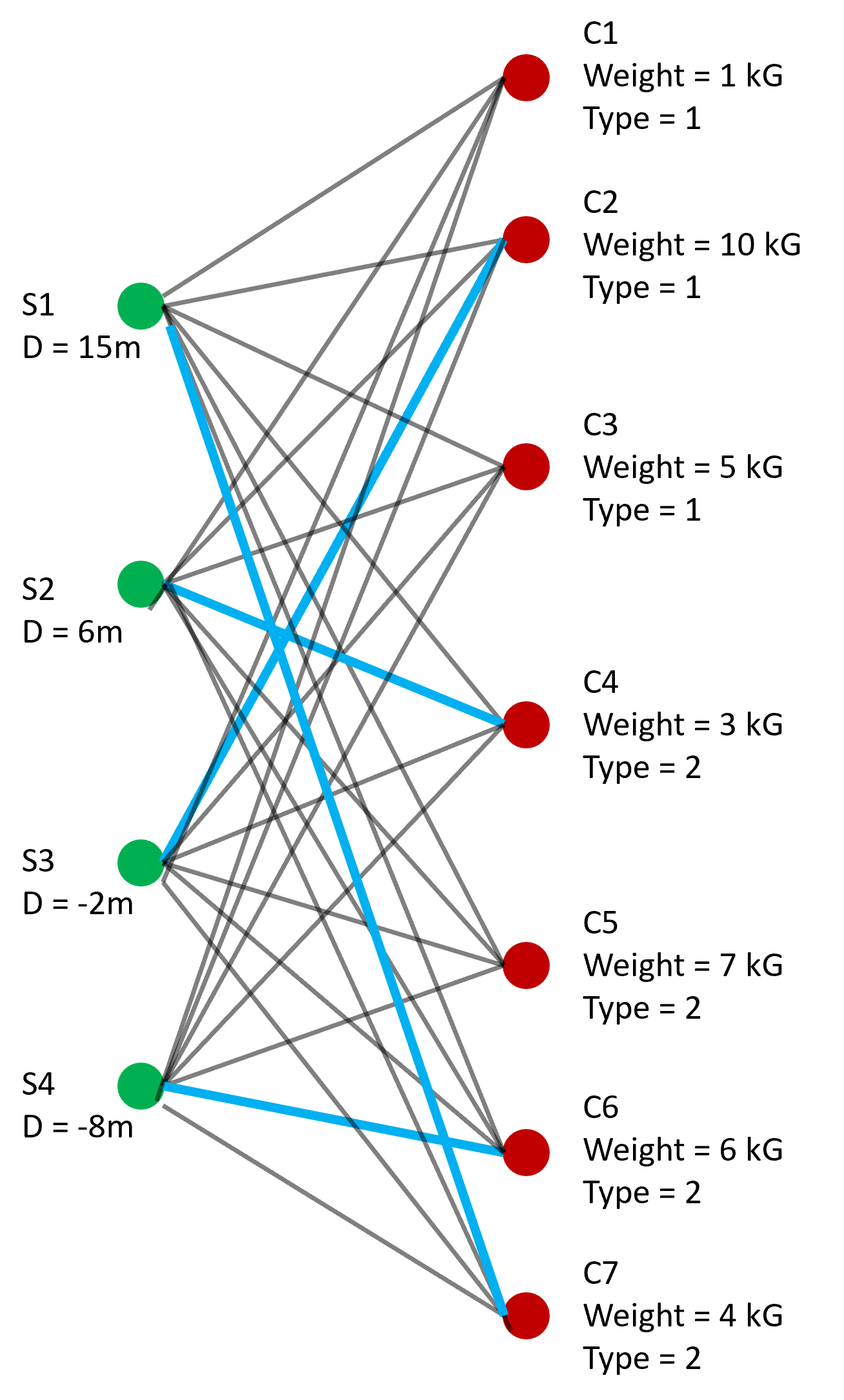}
\hfill
{b)}\includegraphics[width=0.45\textwidth]{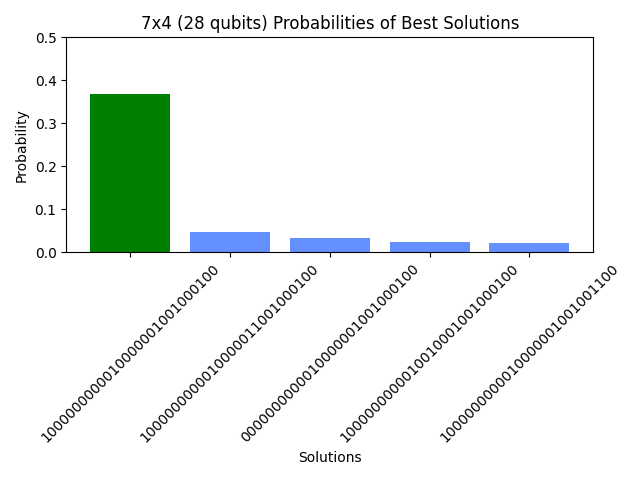}
\caption{Forte QPU results for 7 containers, 4 slots (28 qubits). Maximum loading weight = 23kG. a) The optimal solution for container assignment, b) Probability histogram of results - the green bar corresponds to the optimal solution. }
\label{fig:forte_results_7-4_2}
\end{figure}

In addition to the full hybrid optimization experiment on the Forte QPU, the same problem instance with 7 containers, 4 slots was also executed on the Forte QPU as an ``infererence'' experiment. The problem was first optimized on the internal IonQ Forte noisy quantum simulator. Ten independent instances of the problem were optimized starting with random initial parameters. The maximum weight constraint was set at 23kG and the quantum circuit had 28 qubits, 56 one qubit gates and 56 two qubit gates. The runs converged to different solutions, some of which were optimal. The solution which corresponds to the Forte QPU hybrid optimization experiment is shown in Fig~\ref{fig:forte_optimization_7-4_results}. Another solution, different from the one shown in Fig~\ref{fig:forte_optimization_7-4_results} but which satisfies all the constraints and has the same objective function value as the solution in Fig~\ref{fig:forte_optimization_7-4_results} was chosen to be executed as an ``inference'' experiment on Forte. The final optimal parameters for this new solution were determined. The quantum circuit initialized with these optimal parameters was then run on the Forte QPU and the output probability histogram was computed using 10000 shots for sampling. The probability histograms as well as the optimal solution sampled for the problem instance are shown in Fig~\ref{fig:forte_results_7-4_2}. Only the top 5 highest probability solutions are plotted in the histograms. The results show the optimal solution 
%found in each case 
with a green bar.

\begin{comment}
\begin{figure}[H]
\centering
{a)}\includegraphics[width=0.4\textwidth]{Figs/forte_results_1.png}
\hfill
{b)}\includegraphics[width=0.4\textwidth]{Figs/forte_results_2.png}
\hfill
{c)}\includegraphics[width=0.3\textwidth]{Figs/forte_results_3.png}
\caption{Forte QPU results for 7 containers, 4 slots (28 qubits). Maximum loading weight = 23kG. Probability histogram of results - the green bars corresponds to the valid solutions of different total weights. Results from two different runs are shown in a) and b). The optimal solution is one with total weight = 23kG. c) The optimal solution for container assignment for best result in (a). }
\label{fig:forte_results_7-4}
\end{figure}
\end{comment}

The results show that the optimal solution is the state with the highest probability. The probability of the optimal solution is $\approx$ 35\% and is once again, well separated from the other sampled solutions showing that the measured probability distribution is not affected greatly by QPU noise. This demonstrates that the Forte QPU has high 2 qubit gate fidelities and low State Preparation and Measurement (SPAM) errors. This experiment also shows the capability of the algorithm to converge to different degenerate solutions (solutions with the same objective value). This capability will be very relevant  when scaling this problem to very large sizes since the number of potential degenerate solutions (not necessarily optimal but near-optimal solutions) increases with increasing number of variables. Successfully converging to any one of these degenerate solutions will provide a path for the quantum algorithm to be an effective and industrially relevant option for solving the aircraft loading problem.

\section{Conclusions and Outlook}\label{conclusion}

The problem of finding the optimal loading of cargo containers onto an aircraft subject to operational constraints is important to airlines for maximizing revenue-generating aircraft payload while minimizing fuel burn and achieving sustainability targets. Here, we studied formulations of the problem that are suitable for a treatment on quantum computers. We showed that, in a QUBO formulation, the problem can be solved effectively using the Multi-Angle Layered Variational Quantum Algorithm (MAL-VQA) to identify the optimal solution on IonQ trapped ion hardware. In a proof of principle demonstration we found the optimal solution for choosing from up to 7 containers of 3 different size types for loading into 4 cargo slots. This requires the use of 28 (number of slots times number of containers) algorithmic qubits \cite{IonQ_blog, lubinski2023application}.
%To be of practical value, the ability to handle more containers, slots, and constraints will be required. In the near future, the number of available algorithmic qubits will exceed what can be simulated on a classical computer. At this point in time, quantum advantage for optimization problems such as cargo loading might become a distinct possibility.

To be of practical value the ability to solve for many more containers and slots will be required. This increase in scale could come into fruition by leveraging multiple ideas, or combinations thereof: first, with the underlying optimization problem being of knapsack type, an interesting direction for future research is to investigate decomposition methods that allow to breaking down a large instance of the problem into smaller, more manageable subproblems. These are then solved using specialized techniques which can include a combination of quantum and classical methods. In the context of the Graph Partitioning Problem which arises e.g. in the context of load balancing, a similar strategy has led to results that are competitive with even very large scale solvers~\cite{lsdyna:2025}. Next, besides the MAL-VAQ approach studied here, there are many other quantum approaches to tackle the aircraft loading problem that can be studied and that have found application in other areas. An example is the Quantum Imaginary Time Evolution quantum algorithm which was applied for instance to MaxCut and other combinatorial optimization problems~\cite{varqite:2024}. Finally, to treat problems at larger scale, also the prospect of  increasing size of quantum hardware is an important factor. As in the future it is expected that hardware will advance in this direction, further investigations will allow to test the proposed algorithm for larger problem sizes in order to assess its prospects for commercial value.

\bibliographystyle{plain}
\bibliography{references}

\begin{thebibliography}{10}

\bibitem{IonQ_blog}
Ionq, algorithmic qubits: A better single number metric.
\newblock \url{https://ionq.com/resources/ algorithmic-qubits-a-better-single-number-metric}.

\bibitem{IonQ_Aria}
Ionq aria.
\newblock \url{https://ionq.com/quantum-systems/aria}.

\bibitem{IonQ_Forte}
Ionq forte.
\newblock \url{https://ionq.com/news/ionq-announces-fourth-quarter-and-full-year-2023-financial-results}.

\bibitem{lsdyna:2025}
Willie Aboumrad, Daiwei Zhu, Claudio Girotto, Fran\c cois-Henry Rouet, Jezer Jojo, Robert Lucas, Jay Pathak, Ananth Kaushik, and Martin Roetteler.
\newblock Accelerating large-scale linear algebra using variational quantum imaginary time evolution.
\newblock {\em arXiv preprint arXiv:2503.13128}, 2025.

\bibitem{Barkoutsos2020improving}
Panagiotis~Kl. Barkoutsos, Giacomo Nannicini, Anton Robert, Ivano Tavernelli, and Stefan Woerner.
\newblock Improving {V}ariational {Q}uantum {O}ptimization using {CV}a{R}.
\newblock {\em {Quantum}}, 4:256, April 2020.

\bibitem{Herrman2022multi-angle}
Rebekah Herrman, Phillip Lotshaw, James Ostrowski, Travis Humble, and Siopsis George.
\newblock Multi-angle quantum approximate optimization algorithm.
\newblock {\em Scientific Reports}, 12, 2022.

\bibitem{lubinski2023application}
Thomas Lubinski, Sonika Johri, Paul Varosy, Jeremiah Coleman, Luning Zhao, Jason Necaise, Charles~H Baldwin, Karl Mayer, and Timothy Proctor.
\newblock Application-oriented performance benchmarks for quantum computing.
\newblock {\em IEEE Transactions on Quantum Engineering}, 2023.

\bibitem{maksymov2023enhancing}
Andrii Maksymov, Jason Nguyen, Yunseong Nam, and Igor Markov.
\newblock Enhancing quantum computer performance via symmetrization.
\newblock {\em arXiv preprint arXiv:2301.07233}, 2023.

\bibitem{varqite:2024}
{Titus D.} Morris, Ananth Kaushik, Martin Roetteler, and {Phillip C.} Lotshaw.
\newblock Performant near-term quantum combinatorial optimization.
\newblock {\em arXiv preprint arXiv:2404.16135}, 2024.

\bibitem{sorensen1999quantum}
Anders S{\o}rensen and Klaus M{\o}lmer.
\newblock Quantum computation with ions in thermal motion.
\newblock {\em Physical review letters}, 82(9):1971, 1999.

\end{thebibliography}

\end{document}